\documentclass[aps,prd,showpacs,floatfix,eqsecnum,amsmath,amssymb,
nofootinbib,twocolumn, showpacs
]{revtex4-1}

\usepackage{graphicx}
\usepackage{dcolumn}
\usepackage{bm}
\usepackage{hyperref}
\usepackage{color}
\begin{document}

\title{%
  Constraint Propagation of $C^2$-adjusted Formulation II\\
  --- Another Recipe for Robust Baumgarte-Shapiro-Shibata-Nakamura \\
  Evolution System ---
}%

\author{Takuya Tsuchiya}
\email{tsuchiya@akane.waseda.jp}
\author{Gen Yoneda}%
\affiliation{%
  Department of Mathematical Sciences, Waseda University, Okubo, Shinjuku,
  Tokyo, 169-8555, Japan
}%
\author{Hisa-aki Shinkai}
\affiliation{%
  Faculty of Information Science and Technology, Osaka Institute of
  Technology, 1-79-1 Kitayama, Hirakata, Osaka 573-0196, Japan\\
  Computational Astrophysics Laboratory,
  Institute of Physical \& Chemical Research (RIKEN),
  Hirosawa, Wako, Saitama, 351-0198 Japan
}%

\date{\today}

\begin{abstract}
  In order to obtain an evolution system which is robust against the 
  violation of constraints, we present a new set of evolution systems based 
  on the so-called Baumgarte-Shapiro-Shibata-Nakamura (BSSN) equations.
  The idea is to add functional derivatives of the norm of constraints,
  $C^2$, to the evolution equations, which was proposed by Fiske (2004) and
  was applied to the ADM formulation in our previous study.
  We derive the constraint propagation equations, discuss the behavior of
  constraint damping, and present the results of numerical tests using the 
  gauge-wave and polarized Gowdy wave spacetimes.
  The construction of the $C^2$-adjusted system is straightforward.  
  However, in BSSN, there are two kinetic constraints and three algebraic 
  constraints; thus, the definition of $C^2$ is a matter of concern.  
  By analyzing constraint propagation equations, we conclude that $C^2$ 
  should include all the constraints, which is also confirmed numerically.
  By tuning the parameters, the lifetime of the simulations can be increased
  as 2-10 times as longer than those of the standard BSSN evolutions.
\end{abstract}
\pacs{04.25.D-}

\maketitle

\section{Introduction}
\label{introduction}

When solving the Einstein equations numerically, the standard way is to
split the spacetime into space and time.
The most fundamental decomposition of the Einstein equations is the 
Arnowitt-Deser-Misner (ADM) formulation \cite{ADM62, York78}.
However, it is well known that in long-term evolutions in strong 
gravitational fields such as the coalescences of binary neutron stars and/or
black holes, simulations with the ADM formulation are unstable and are often
interrupted before producing physically interesting results.
Finding more robust and stable formulations is known to the ``formulation 
problem'' in numerical relativity \cite{SY00,Shinkai09,SY02gr-qc}.

Many formulations have been proposed in the last two decades.
The most commonly used sets of evolution equations among numerical 
relativists are the so-called Baumgarte-Shapiro-Shibata-Nakamura (BSSN) 
formulation \cite{SN95,BS98}, the generalized harmonic (GH) formulation 
\cite{PretriusCQG05, Garfinkle02}, the Kidder-Scheel-Teukolsky (KST) 
formulation \cite{KST01}, and the Z4 formulation \cite{BLPZ03,BLPZ04} 
(as references of their numerical application, we here cite only well-known 
articles; \cite{BCCKM06, CLMZ06} for the BSSN formulation, 
\cite{Pretorius05} for the GH formulation, \cite{SBCKMP09} for the KST 
formulation, and \cite{ABBRP11} for the Z4 formulation).

All of the above modern formulations include the technique of 
``constraint damping'', which attempts to control the violations of 
constraints by adding the constraint terms to their evolution equations.
Using this technique, more stable and accurate systems are obtained 
(see e.g. \cite{WBH11,GCHM05}).
This technique can be described as `adjustment' of the original system.

In \cite{YS01prd, YS02, SY02}, two of the authors systematically 
investigated how the adjusted terms change the original systems by 
calculating the constraint propagation equations.
The authors suggested some effective adjustments for the BSSN formulation
under the name ``adjusted BSSN formulation''\cite{YS02}.
The actual constraint-damping effect was confirmed by numerical tests 
\cite{KS08}.

Fiske proposed a method of adjusting the original evolution system using the
norm of the constraints, $C^2$, \cite{Fiske04}, which we call a 
``$C^2$-adjusted system.''
The new evolution equations force the constraints to evolve towards their 
decay if the coefficient parameters of the adjusted terms are set as 
appropriate positive values.
Fiske reported the damping effect of the constraint violations for the 
Maxwell system \cite{Fiske04} and for the linearized ADM and BSSN 
formulations \cite{Fiske_Phd}.
He also reported the limitation of the magnitude of the coefficient 
parameters of the adjusted terms.

In \cite{TYS11}, we applied this $C^2$-adjusted system to the (full) ADM 
formulation and presented some numerical tests. 
We confirmed that the violations of the constraints are less than those in 
the original system.
We also reported the differences of the effective range of the coefficient 
of the adjusted terms.

In this article, we apply the $C^2$-adjusted system to the (full) BSSN 
formulation and derive the constraint propagation equations in the flat 
space.
We perform some numerical tests and compare them with three other types of 
BSSN formulations: the standard BSSN formulation, the 
$\widetilde{A}$-adjusted BSSN formulation, and the $C^2$-adjusted BSSN 
formulation.
We use the gauge-wave and polarized Gowdy wave testbeds, which are the test
problems as is known to apples-with-apples testbeds for comparing evolution 
systems \cite{Alcubierre_etc04}.
Since the models are precisely fixed up to the gauge conditions, boundary 
conditions, and technical parameters, the testbeds are widely used for
comparisons \cite{KS08, Zumbusch09, BB10}.

The structure of this article is as follows.
We review the ideas of adjusted systems and $C^2$-adjusted system in 
Sec.\ref{GeneralIdea}.
In Sec.\ref{ApplicationEinsteinEq}, we review the standard and adjusted BSSN
formulations and derive the $C^2$-adjusted version of the BSSN formulation.
In Sec.\ref{NumericalExamples}, we present some numerical tests of the
gauge-wave and polarized Gowdy wave testbeds. 
We show the damping effect of the constraint violations, and confirm that 
inclusion of algebraic constraints in $C^2$ make the violations of 
constraints decrease.
We summarize this article in Sec.\ref{Summary}.
In this article, we only consider vacuum spacetime, but the inclusion of
matter is straightforward.


\section{Ideas of adjusted systems and $C^2$-adjusted systems}
\label{GeneralIdea}

\subsection{Idea of adjusted systems}
\label{GeneralIdea_AdjustedSystems}

Suppose we have dynamical variables $u^i$ that evolve with the evolution 
equations
\begin{align}
  &\partial_t u^i = f(u^i, \partial_j u^i, \cdots),
  \label{eq:generalEvolveEquations}
\end{align}
and suppose also that the system has the (first class) constraint equations
\begin{align}
  &C^a(u^i, \partial_j u^i, \cdots)\approx 0.
  \label{eq:generalConstraintEquations}
\end{align}
We can then predict how the constraints are preserved by evaluating the 
constraint propagation equations
\begin{align}
  \partial_t C^a &=g(C^a,\partial_i C^a,\cdots),
  \label{eq:generalConstraintPropagation}
\end{align}
which measure the violation behavior of constraints $C^a$ in time evolution.
Equation \eqref{eq:generalConstraintPropagation} is theoretically weakly
zero, i.e., $\partial_t C^a \approx 0$, since the system is supposed to be 
the first class. 
However, free numerical evolution with discretized grids introduces a 
constraint violation, at least at the level of truncation error, which 
sometimes grows and stops the simulations.
The unstable feature of ADM evolution can be understood on the basis of this
analysis \cite{Pretorius05}.

Such features of the constraint propagation equations,
\eqref{eq:generalConstraintPropagation}, change when we modify the original
evolution equations.
Suppose we add constraint terms to the right-hand-side of
\eqref{eq:generalEvolveEquations} as
\begin{align}
  \partial_t u^i = f(u^i, \partial_j u^i, \cdots) + F(C^a, \partial_j C^a, 
  \cdots),
  \label{eq:generalADjustedEvolutionEqs}
\end{align}
where $F(C^a,\cdots)\approx0$ in principle zero but not exactly zero in 
numerical evolutions.
With this adjustment, equation \eqref{eq:generalConstraintPropagation} will 
also be modified to
\begin{align}
  \partial_t C^a&=g(C^a,\partial_i C^a,\cdots) + G(C^a, \partial_i C^a,
  \cdots).
  \label{eq:generalADjustedConstraintPropagationEqs}
\end{align}
Therefore, we are able to control $\partial_t C^a$ by making an appropriate 
adjustment $F(C^a, \partial_j C^a, \cdots)$ in 
\eqref{eq:generalADjustedEvolutionEqs}.
If $\partial_t C^a<0$ is realized, then the system has the constraint 
surface as an attractor.

This technique is also known as a constraint-damping technique.
Almost all the current popular formulations used in large-scale numerical 
simulations include this implementation.
The purpose of this article is to find a better way of adjusting the 
evolution equations to realize $\partial_t C^a\leq 0$.


\subsection{Idea of $C^2$-adjusted systems}
\label{theideaofC2}

Fiske \cite{Fiske04} proposed a way of adjusting the evolution equations
which we call ``$C^2$-adjusted systems'';
\begin{align}
  \partial_t u^i = f(u^i, \partial_j u^i, \cdots)-\kappa^{i j}
  \left(\frac{\delta C^2}{\delta u^j}\right),
  \label{eq:adjutedGeneralEvolveEquations}
\end{align}
where $\kappa^{i j}$ is a positive-definite constant coefficient and $C^2$ 
is the norm of the constraints, which is defined as 
$\displaystyle{C^2\equiv \int C_a C^a d^3 x}$.
The term $(\delta C^2/\delta u^j)$ is the functional derivative of $C^2$ 
with respect to $u^j$.
The associated constraint propagation equation becomes
\begin{align}
  \partial_t C^2 = h(C^a, \partial_i C^a, \cdots) - \int d^3 x
  \left(\frac{\delta C^2}{\delta u^i}\right)\kappa^{i j}
  \left(\frac{\delta C^2}{\delta u^j}\right).
  \label{eq:adjustedGeneralConstraintPropagation_of_C2}
\end{align}

The motivation for this adjustment is to naturally obtain the 
constraint-damping system, $\partial_t C^2<0$.
If we set $\kappa^{i j}$ so that the second term of the right-hand side of
\eqref{eq:adjustedGeneralConstraintPropagation_of_C2} becomes larger than
the first term, then $\partial_t C^2$ becomes negative, which indicates that
constraint violations are expected to decay to zero.
Fiske presented numerical examples of the Maxwell system and the linearized
ADM and BSSN formulations, and concluded that this method actually reduces 
constraint violations as expected.
In our previous work \cite{TYS11}, we applied the $C^2$-adjusted system to 
the (full) ADM formulation and derived the constraint propagation 
equations. 
We confirmed that $\partial_t C^2<0$ is expected in the flat spacetime.
We performed numerical tests with the $C^2$-adjusted ADM formulation using
the Gowdy wave testbed, and confirmed that the violations of the constraint
are lower than those of the standard ADM formulation.
The simulation continues 1.7 times longer than that of the standard ADM 
formulation with the magnitude of the violations of the constraint less than
order $O(10^0)$.



\section{Application to BSSN formulation}
\label{ApplicationEinsteinEq}

\subsection{Standard BSSN Formulation}
\label{standardBSSN}

We work with the widely used notation of the BSSN system.
That is, the dynamical variables 
$(\varphi, K, \widetilde{\gamma}_{i j}, \widetilde{A}_{i j}, 
\widetilde{\Gamma}^i)$ as the replacement of the variables of the ADM 
formulation, $(\gamma_{i j}, K_{i j})$, where
\begin{align}
  \varphi
  & \equiv (1/12) \log({\rm det} (\gamma_{i j})),
  \label{eq:phi_BSSNVariables}\\
  K
  & \equiv \gamma^{i j} K_{i j},
  \label{eq:K_BSSNVariables}\\
  \widetilde{\gamma}_{i j}
  & \equiv  e ^{-4 \varphi} \gamma_{i j},
  \label{eq:gammaTilde_BSSNVariables}\\
  \widetilde{A}_{i j}
  & \equiv  e ^{-4 \varphi}(K_{i j} - (1/3) \gamma_{i j} K), \,\,{\rm and}
  \label{eq:A_BSSNVariables}\\
  \widetilde{\Gamma}^i
  & \equiv \widetilde{\gamma}^{m n} \widetilde{\Gamma}^i{}_{m n}.
  \label{eq:Gamma_BSSNVariables}
\end{align}
The BSSN evolution equations are, then,
\begin{align}
  \partial_t \varphi
  & = - (1/6) \alpha K + (1/6) (\partial_i \beta^i) + \beta^i
  (\partial_i \varphi),
  \label{eq:phi_standardBSSNEvolutionEquations}\\
  \partial_t K
  & = \alpha \widetilde{A}_{i j} \widetilde{A}^{i j} + (1/3) \alpha K^2
  - D_i D^i \alpha + \beta^i (\partial_i K),
  \label{eq:K_standardBSSNEvolutionEquations}\\
  \partial_t\widetilde{\gamma}_{i j}
  & = -2 \alpha \widetilde{A}_{i j} - (2/3) \widetilde{\gamma}_{i j}
  (\partial_\ell \beta^\ell)
  \nonumber\\
  &\quad
  + \widetilde{\gamma}_{j \ell} (\partial_i \beta^\ell)
  + \widetilde{\gamma}_{i \ell} (\partial_j \beta^\ell)
  + \beta^\ell (\partial_\ell \widetilde{\gamma}_{i j}),
  \label{eq:gamma_standardBSSNEvolutionEquations}\\
  \partial_t \widetilde{A}_{i j}
  & = \alpha K \widetilde{A}_{i j}
  - 2 \alpha \widetilde{A}_{i \ell} \widetilde{A}^\ell{}_j
  + \alpha  e ^{-4 \varphi} R_{i j}{}^{\rm TF}
  \nonumber\\
  &\quad
  - e ^{-4\varphi}(D_i D_j\alpha)^{\rm TF}
  - (2/3) \widetilde{A}_{i j}(\partial_\ell \beta^\ell)
  \nonumber\\
  &\quad
  + (\partial_i \beta^\ell) \widetilde{A}_{j \ell}
  + (\partial_j \beta^\ell) \widetilde{A}_{i \ell}
  + \beta^\ell (\partial_\ell \widetilde{A}_{i j}),
  \label{eq:A_standardBSSNEvolutionEquations}\\
  \partial_t \widetilde{\Gamma}^i
  & = 
  2 \alpha\{ 6 (\partial_j \varphi) \widetilde{A}^{i j}
  + \widetilde{\Gamma}^i{}_{j \ell} \widetilde{A}^{j \ell}
  - (2/3) \widetilde{\gamma}^{i j} (\partial_j K)
  \}
  \nonumber\\
  &\quad
  -2 (\partial_j \alpha) \widetilde{A}^{i j}
  + (2/3) \widetilde{\Gamma}^i (\partial_j \beta^j)
  + (1/3) \widetilde{\gamma}^{i j} (\partial_\ell \partial_j \beta^\ell)
  \nonumber\\
  &\quad
  + \beta^\ell (\partial_\ell \widetilde{\Gamma}^i)
  - \widetilde{\Gamma}^j (\partial_j \beta^i)
  + \widetilde{\gamma}^{j \ell} (\partial_j \partial_\ell \beta^i),
  \label{eq:CGamma_standardBSSNEvolutionEquations}
\end{align}
where ${}^{\rm TF}$ denotes the trace-free part. The Ricci tensor in the 
BSSN system is normally calculated as
\begin{align}
  R_{i j}
  & \equiv \widetilde{R}_{i j} + R^\varphi_{i j},
  \label{eq:ricci_standardBSSNFormulation}
\end{align}
where
\begin{align}
  \widetilde{R}_{i j}
  &\equiv \widetilde{\gamma}_{n (i} \partial_{j)} \widetilde{\Gamma}^n
  + \widetilde{\gamma}^{\ell m} (2 \widetilde{\Gamma}^k{}_{\ell (i}
  \widetilde{\Gamma}_{j) k m} + \widetilde{\Gamma}_{n \ell j}
  \widetilde{\Gamma}^n{}_{i m})
  \nonumber\\
  &\quad
  - (1/2) \widetilde{\gamma}^{m \ell} \widetilde{\gamma}_{i j, m \ell}
  + \widetilde{\Gamma}^n \widetilde{\Gamma}{}_{(i j) n},
  \label{eq:ricciTilde_standardBSSNFormulation}\\
  R_{i j}^\varphi
  & \equiv - 2 \widetilde{D}_i \widetilde{D}_j \varphi
  + 4 (\widetilde{D}_i \varphi) (\widetilde{D}_j \varphi)
  - 2 \widetilde{\gamma}_{i j} \widetilde{D}_m \widetilde{D}^m \varphi
  \nonumber\\
  &\quad
  -4 \widetilde{\gamma}_{i j} (\widetilde{D}^m \varphi)
  (\widetilde{D}_m \varphi).
  \label{eq:ricciPhi_standardBSSNFormulation}
\end{align}

The BSSN system has five constraint equations.
The ``kinematic'' constraint equations, which are the Hamiltonian constraint
equation and the momentum constraint equations ($\mathcal{H}$-constraint and
$\mathcal{M}$-constraint, hereafter), are expressed in terms of the BSSN
basic variables as
\begin{align}
  \mathcal{H}
  & \equiv  e ^{-4 \varphi} \widetilde{R} - 8  e ^{-4 \varphi}
  (\widetilde{D}_i \widetilde{D}^i \varphi + (\widetilde{D}^m \varphi)
  (\widetilde{D}_m \varphi))
  \nonumber\\
  &\quad
  + (2/3)K^2 - \widetilde{A}_{i j} \widetilde{A}^{i j} -(2/3) \mathcal{A} K
  \approx 0
  \label{eq:hamiltonian_standardBSSNConstraintEquations},\\
  \mathcal{M}_i
  & \equiv -(2/3) \widetilde{D}_i K + 6 (\widetilde{D}_j \varphi)
  \widetilde{A}^j{}_i + \widetilde{D}_j \widetilde{A}^j{}_i
  \nonumber\\
  &\quad
  -2(\widetilde{D}_i \varphi) \mathcal{A}\approx 0,
  \label{eq:momentum_standardBSSNConstraintEquations}
\end{align}
respectively, where $\widetilde{D}_i$ is the covariant derivative associated
with $\widetilde{\gamma}_{i j}$ and $\widetilde{R}=\widetilde{\gamma}^{i j}
\widetilde{R}_{i j}$.
Because of the introduction of new variables, there are additional 
``algebraic'' constraint equations:
\begin{align}
  \mathcal{G}^i
  & \equiv \widetilde{\Gamma}^i - \widetilde{\gamma}^{j \ell}
  \widetilde{\Gamma}^i{}_{j \ell}\approx 0,
  \label{eq:gConstraint_standardBSSNConstraintEquations}\\
  \mathcal{A}
  & \equiv \widetilde{A}^{i j} \widetilde{\gamma}_{i j}\approx 0,
  \label{eq:aCosntraint_standardBSSNConstraintEquations}\\
  \mathcal{S}
  & \equiv{\rm det} (\widetilde{\gamma}_{i j}) - 1\approx 0,
  \label{eq:sConstraint_standardBSSNConstraintEquations}
\end{align}
which we call the $\mathcal{G}$-, $\mathcal{A}$-, and 
$\mathcal{S}$-constraints, respectively, hereafter.
If the algebraic constraint equations,
\eqref{eq:gConstraint_standardBSSNConstraintEquations}-\eqref{eq:sConstraint_standardBSSNConstraintEquations},
are not satisfied, the BSSN formulation and ADM formulation are not
equivalent mathematically.


\subsection{$C^2$-adjusted BSSN Formulation}
\label{c2adjustedBSSN}

The $C^2$-adjusted BSSN evolution equations are formally written as
\begin{align}
  \partial_t \varphi
  &= \eqref{eq:phi_standardBSSNEvolutionEquations}
  - \lambda_\varphi \left(\frac{\delta C^2}{\delta \varphi}\right),
  \label{eq:phi_c2adjusted_BSSN}\\
  \partial_t K
  &= \eqref{eq:K_standardBSSNEvolutionEquations}
  - \lambda_K \left(\frac{\delta C^2}{\delta K}\right),
  \label{eq:K_c2adjusted_BSSN}\\
  \partial_t \widetilde{\gamma}_{i j}
  &= \eqref{eq:gamma_standardBSSNEvolutionEquations}
  - \lambda_{\widetilde{\gamma} i j m n}
  \left(\frac{\delta C^2}{\delta \widetilde{\gamma}_{m n}}\right),
  \label{eq:gamma_c2adjusted_BSSN}\\
  \partial_t \widetilde{A}_{i j}
  &= \eqref{eq:A_standardBSSNEvolutionEquations}
  - \lambda_{\widetilde{A} i j m n}
  \left(\frac{\delta C^2}{\delta \widetilde{A}_{m n}}\right),
  \label{eq:A_c2adjusted_BSSN}\\
  \partial_t \widetilde{\Gamma}^i
  &= \eqref{eq:CGamma_standardBSSNEvolutionEquations}
  - \lambda_{\widetilde{\Gamma}}^{i j}
  \left(\frac{\delta C^2}{\delta \widetilde{\Gamma}^j}\right),
  \label{eq:CGamma_c2adjusted_BSSN}
\end{align}
where all the coefficients $\lambda_\varphi$, $\lambda_K$, 
$\lambda_{\widetilde{\gamma}}{}_{i j m n}$, 
$\lambda_{\widetilde{A}}{}_{i j m n}$, and 
$\lambda_{\widetilde{\Gamma}}^{i j}$ are positive definite.
$C^2$ is a function of the constraints $\mathcal{H}$, $\mathcal{M}_i$,
$\mathcal{G}^i$, $\mathcal{A}$, and $\mathcal{S}$, which we set as 
\begin{align}
  C^2 
  &= \int\Large(\mathcal{H}^2 + \gamma^{i j}\mathcal{M}_i \mathcal{M}_j
  + c_G\gamma_{i j}\mathcal{G}^i\mathcal{G}^j \nonumber\\
  &\qquad+ c_A\mathcal{A}^2 + c_S\mathcal{S}^2\Large) d^3x,
  \label{eq:c2Definition_tmp}
\end{align}
where, $c_G$, $c_A$, and $c_S$ are Boolean parameters (0 or 1).
These three parameters are introduced to prove the necessity of the 
algebraic constraint terms in \eqref{eq:c2Definition_tmp}.

The adjusted terms in 
\eqref{eq:phi_c2adjusted_BSSN}-\eqref{eq:CGamma_c2adjusted_BSSN} are then
written down explicitly, as shown in Appendix \ref{Appendix_adjustTermsC2}.
The constraint propagation equations of this system are also derived for
the Minkowskii background, as shown in Appendix 
\ref{Appendix_CP_Minkowskii}.

Now we discuss the effect of the algebraic constraints.
From \eqref{eq:CPH_Minkowskii}-\eqref{eq:CPS_Minkowskii}, we see that the
constraints affect each others.
The constraint propagation equations of the algebraic constraints, 
\eqref{eq:CPG_Minkowskii}-\eqref{eq:CPS_Minkowskii}, include 
$c_G(\lambda_{\widetilde{\gamma}}\Delta\delta^a{}_b
-2\lambda_{\widetilde{\Gamma}}\delta^a{}_b)\mathcal{G}^b$,
$-6c_{A} \lambda_{\widetilde{A}}\mathcal{A}$, and 
$-6c_S\lambda_{\widetilde{\gamma}}\mathcal{S}$, respectively.
These terms contribute to reduce the violations of each constraint if
$c_G$, $c_A$, and $c_S$ are non-zero.
Therefore, we adopt $c_G=c_A=c_S=1$ in \eqref{eq:c2Definition_tmp};
\begin{align}
  C^2 = \int \left(\mathcal{H}^2 + \gamma^{i j}\mathcal{M}_i \mathcal{M}_j
  + \gamma_{i j}\mathcal{G}^i\mathcal{G}^j + \mathcal{A}^2
  +\mathcal{S}^2\right)d^3x.
  \label{eq:C2definition}
\end{align}
This discussion is considered only from the viewpoint of the inclusion of 
the diffusion terms.
In order to validate this decision, we perform some numerical examples
in Sec.\ref{NumericalExamples}.


\subsection{$\widetilde{A}$-adjusted BSSN System}
\label{adjustedSystem}

In \cite{YS02}, two of the authors reported some examples of adjusted
systems for the BSSN formulation.
The authors investigated the signatures of eigenvalues of the coefficient 
matrix of the constraint propagation equations, and concluded 
three of the examples to be the best candidates for the adjustment.
The actual numerical tests were performed later \cite{KS08} using the 
gauge-wave, linear-wave, and polarized Gowdy wave testbeds.
The most robust system among the three examples for these three testbeds was
the $\widetilde{A}$-adjusted BSSN formulation, which replaces 
\eqref{eq:A_standardBSSNEvolutionEquations} in the standard BSSN system with
\begin{align}
  \partial_t\widetilde{A}_{i j}
  &=\eqref{eq:A_standardBSSNEvolutionEquations} +
  \kappa_A\alpha \widetilde{D}_{(i}\mathcal{M}_{j)},
  \label{eq:A-equation}
\end{align}
where $\kappa_A$ is a constant.
If $\kappa_A$ is set as positive, the violations of the constraints are 
expected to be damped in flat spacetime \cite{YS02}.
We also use the $\widetilde{A}$-adjusted BSSN system for comparison in the 
following numerical tests. 



\section{Numerical Examples}
\label{NumericalExamples}

\begin{table*}[t]
  \caption{List of figures.\label{table:figures}}
  \begin{tabular}{llll}
    \hline
    & & gauge-wave test & Gowdy wave test \\
    & & \S \ref{gauge-wave_testbed} & \S \ref{GOWDY_TEST}\\
    \hline
    (A) & standard BSSN 
    \eqref{eq:phi_standardBSSNEvolutionEquations}-\eqref{eq:CGamma_standardBSSNEvolutionEquations}
    & Fig.\ref{fig:ConstraintViolations_GaugeWave} norm each 
    & Fig.\ref{fig:ConstraintViolationsGowdy} norm each\\
    & (constraint propagation, see App. \ref{Appendix_CP_noshit})
    & Fig.\ref{fig:C2_GaugeWave} norm all
    & Fig.\ref{fig:DampingViolations_Gowdy} norm all\\
    \hline
    (B) & $\widetilde{A}$-adjusted BSSN
    & Fig.\ref{fig:C2_GaugeWave} norm all 
    & Fig.\ref{fig:DampingViolations_Gowdy} norm all\\
    & \eqref{eq:phi_standardBSSNEvolutionEquations}-\eqref{eq:gamma_standardBSSNEvolutionEquations},
\eqref{eq:CGamma_standardBSSNEvolutionEquations}, and \eqref{eq:A-equation}
 & Fig.\ref{fig:ConstraintViolationsC2_GaugeWave} 
    norm each & \\
    & (constraint propagation, see App. \ref{Appendix_CP_Minkowskii}) & & 
    \\ \hline
    (C) & $C^2$-adjusted BSSN 
    \eqref{eq:phi_c2adjusted_BSSN}-\eqref{eq:CGamma_c2adjusted_BSSN}
    & Fig.\ref{fig:C2_GaugeWave} norm all 
    & Fig.\ref{fig:DampingViolations_Gowdy} norm all \\
    & (constraint propagation, see App. \ref{Appendix_CP_Minkowskii})
    & Fig.\ref{fig:ConstraintViolationsC2_GaugeWave} norm each 
    & Fig.\ref{fig:C2Constraints} norm each\\
    & & Fig.\ref{fig:DiffereneceGauge} adjusted ratio 
    & Fig.\ref{fig:MagnitudeTermsGowdy} adjusted ratio\\
    & & Fig.\ref{fig:C2definition_GaugeWave} \eqref{eq:C2definition} test 
    & Fig.\ref{fig:C2Parameter_GowdyWave2} \eqref{eq:C2definition} test\\
    \hline
  \end{tabular}
\end{table*}
We test the three systems ($C^2$-adjusted BSSN, $\widetilde{A}$-adjusted 
BSSN, and standard BSSN) in numerical evolutions using the gauge-wave 
and polarized Gowdy wave spacetimes, which are the standard tests for 
comparisons of formulations in numerical relativity, and are known as 
apples-with-apples testbeds \cite{Alcubierre_etc04}.
We also performed the linear-wave testbed but the violations of the 
constraint are negligible; thus, we employ only the above two testbeds in 
this article.
These tests have been used by several groups and were reported in the same
manner (e.g., \cite{Zumbusch09, BB10, KS08, PHK08}).

For simplicity, we set the coefficient parameters in
\eqref{eq:gamma_c2adjusted_BSSN}-\eqref{eq:CGamma_c2adjusted_BSSN} to
$\lambda_{\widetilde{\gamma} i j m n} = \lambda_{\widetilde{\gamma}} 
\delta_{i m}\delta_{j n}$,  $\lambda_{\widetilde{A} i j m n} = 
\lambda_{\widetilde{A}} \delta_{i m}\delta_{j n}$, and 
$\lambda_{\widetilde{\Gamma}}^{i j} = \lambda_{\widetilde{\Gamma}}
\delta^{ij}$ with non-negative coefficient constant parameters 
$\lambda_{\widetilde{\gamma}}$, $\lambda_{\widetilde{A}}$, and
$\lambda_{\widetilde{\Gamma}}$.
Our code passes the convergence test with second-order accuracy.
We list the figures in this article in Table \ref{table:figures} for 
reader's convenience.

\subsection{Gauge-wave Testbed}
\label{gauge-wave_testbed}

\subsubsection{Metric and Parameters}
\label{metric_paramaters_gauge}

The metric of the gauge-wave test is
\begin{align}
  ds^2 = -H dt^2 + H dx^2 + dy^2 + dz^2,
\end{align}
where
\begin{align}
  H=1-A\sin(2\pi (x-t)/d),
\end{align}
which describes a sinusoidal gauge wave of amplitude $A$ propagating along 
the $x$-axis.
The nontrivial extrinsic curvature is
\begin{align}
  K_{xx} = -\frac{\pi A}{d}\frac{\cos(\frac{2\pi (x-t)}{d})}
  {\sqrt{1-A\sin\frac{2\pi(x-t)}{d}}}.
\end{align}
Following \cite{Alcubierre_etc04}, we chose the numerical domain and 
parameters as follows:
\begin{itemize}
\item Gauge-wave parameters: $d=1$ and $A=10^{-2}$.
\item Simulation domain: $x\in [-0.5, 0.5]$, $y=z=0$.
\item Grid: $x^n = -0.5+(n-1/2)dx$ with $n=1,\cdots, 100$, where $dx=1/100$.
\item Time step: $dt = 0.25 dx$.
\item Boundary conditions: Periodic boundary condition in $x$-direction
  and planar symmetry in $y$- and $z$-directions.
\item Gauge conditions:
  \begin{align}
    \partial_t \alpha = -\alpha^2 K, \quad \beta^i = 0.
  \end{align}
\item Scheme: second-order iterative Crank-Nicolson.
\end{itemize}


\subsubsection{Constraint Violations and Their Dampings}
\label{ConstraintviolationsGauge}

\begin{figure}[t]
  \begin{center}
    \includegraphics[keepaspectratio=true,width=85mm]{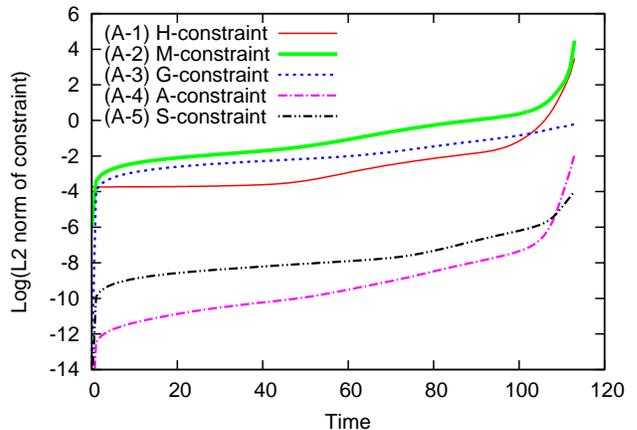}
  \end{center}
  \caption{
    L2 norm of each constraint violation in the gauge-wave evolution using 
    the standard BSSN formulation.
    The vertical axis is the logarithm of the L2 norm of the constraints 
    and the horizontal axis is time.
    We see the evolution stops at $t=110$ due to the growth of 
    $\mathcal{M}$-constraint violation.
    \label{fig:ConstraintViolations_GaugeWave}}
\end{figure}
Figure \ref{fig:ConstraintViolations_GaugeWave} shows the violations of
five constraint equations $\mathcal{H}$, $\mathcal{M}_i$, $\mathcal{G}^i$,
$\mathcal{A}$, and $\mathcal{S}$ for the gauge-wave evolution using the
standard BSSN formulation.
The violation of the $\mathcal{M}$-constraint, line (A-2), is the largest 
during the evolution, while the violations of both the 
$\mathcal{A}$-constraint and $\mathcal{S}$-constraint are negligible.
This is the starting point for improving the BSSN formulation.

\begin{figure}[t]
  \begin{center}
    \includegraphics[keepaspectratio=true,width=85mm]{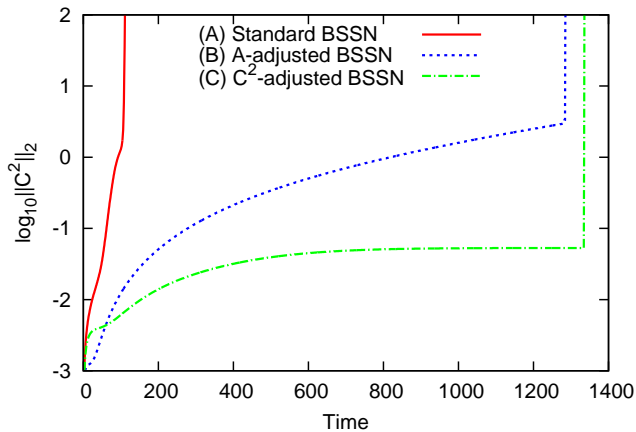}
  \end{center}
  \caption{
    L2 norm of all the constraints in gauge-wave evolution comparing 
    three BSSN formulations:
    (A) standard BSSN formulation (solid line), 
    (B) $\widetilde{A}$-adjusted BSSN formulation (dotted line), and
    (C) $C^2$-adjusted BSSN formulation (dot-dashed line).
    The adopted parameters are $\kappa_A=10^{-1.6}$ for (B), and
    $\lambda_\varphi = 10^{-8.5}$, $\lambda_{K} = 10^{-8.4}$,
    $\lambda_{\widetilde{\gamma}} = 10^{-7.3}$, $\lambda_{\widetilde{A}} =
    10^{-2.5}$, and $\lambda_{\widetilde{\Gamma}} = 10^{-1.8}$ for (C) to 
    minimize $C^2$ at $t=1000$.
    The constraint violations of the $\widetilde{A}$-adjusted BSSN 
    formulation, (B), increase with time and the simulation stops before
    $t=1300$, while those of the $C^2$-adjusted BSSN
    formulation, (C), remain at $O(10^{-1})$ until $t=1300$ and the 
    simulation stops at $t=1350$.
    \label{fig:C2_GaugeWave}}
\end{figure}
Applying the adjustment procedure, the lifetime of the standard BSSN 
evolution is increased at least 10-fold.
In Fig.\ref{fig:C2_GaugeWave}, we plot the L2 norm of the constraints, 
\eqref{eq:C2definition}, of three BSSN evolutions: (A) the standard BSSN 
formulation \eqref{eq:phi_standardBSSNEvolutionEquations}-\eqref{eq:CGamma_standardBSSNEvolutionEquations}, 
(B) the $\widetilde{A}$-adjusted BSSN formulation 
\eqref{eq:phi_standardBSSNEvolutionEquations}-\eqref{eq:gamma_standardBSSNEvolutionEquations},
\eqref{eq:CGamma_standardBSSNEvolutionEquations}, and \eqref{eq:A-equation},
and (C) the $C^2$-adjusted BSSN formulation 
\eqref{eq:phi_c2adjusted_BSSN}-\eqref{eq:CGamma_c2adjusted_BSSN}.
For the standard BSSN case, we see the violation of constraint 
monotonically increases in the earlier stage, while other two adjusted 
cases keep it smaller.
We can say that the $C^2$-adjusted formulation is the most robust one 
against the violation of constraints between three.

\begin{figure*}[t]
  \begin{center}
    \includegraphics[keepaspectratio=true,width=150mm]{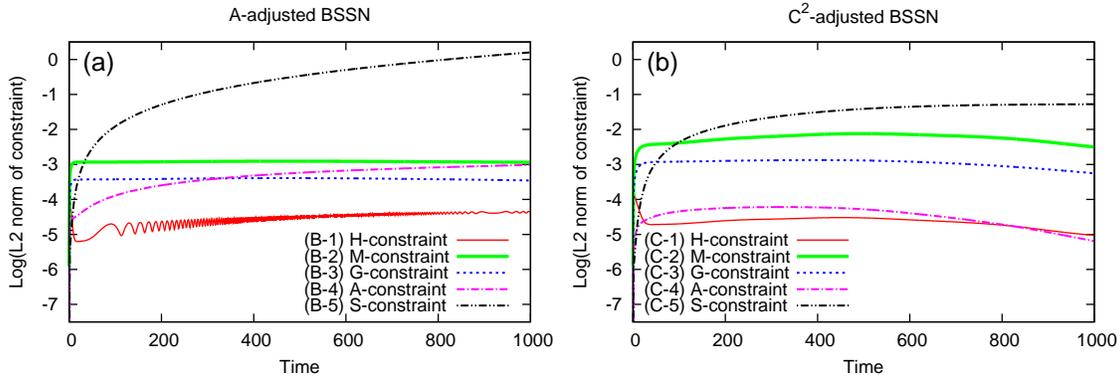}
  \end{center}
  \caption{
    L2 norm of each constraint in the gauge-wave evolution using the
    $\widetilde{A}$-adjusted BSSN formulation [panel (a)] and $C^2$-adjusted
    BSSN formulation [panel (b)].
    The parameters $\kappa_{A}$, $\lambda_\varphi$, $\lambda_{K}$, 
    $\lambda_{\widetilde{\gamma}}$, $\lambda_{\widetilde{A}}$, and
    $\lambda_{\widetilde{\Gamma}}$ are the same as those in 
    Fig.\ref{fig:C2_GaugeWave}.
    In both panels, we see that the violations of the
    $\mathcal{H}$-constraint [the lines (B-1) and (C-1)], the 
    $\mathcal{M}$-constraint [(B-2) and (C-2)], and the 
    $\mathcal{G}$-constraint [(B-3) and (C-3)] are less than those for the 
    standard BSSN formulation in 
    Fig.\ref{fig:ConstraintViolations_GaugeWave}.
    However, the violations of the $\mathcal{A}$-constraint 
    [(B-4) and (C-4)] and the $\mathcal{S}$-constraint [(B-5) and (C-5)] 
    are larger.
    Line (B-5) overlaps with line (B) in Fig.\ref{fig:C2_GaugeWave} after 
    $t=100$, and line (C-5) overlaps with line (C) in 
    Fig.\ref{fig:C2_GaugeWave} after $t=500$.
    \label{fig:ConstraintViolationsC2_GaugeWave}}
\end{figure*}
We plot the norm of each constraint equation in 
Fig.\ref{fig:ConstraintViolationsC2_GaugeWave}.
First, we see that the violation of the $\mathcal{M}$-constraint for the 
two adjusted BSSN formulations [the lines (B-2) and (C-2) in 
Fig.\ref{fig:ConstraintViolationsC2_GaugeWave}] are less than that of the
standard BSSN formulation in Fig.\ref{fig:ConstraintViolations_GaugeWave}.
This behavior would be explained from the constraint propagation equations, 
where we see the terms $\lambda_{\widetilde{A}}\Delta \mathcal{M}_a$ and 
$(1/2)\kappa_A \Delta \mathcal{M}_i$ in \eqref{eq:CPM_Minkowskii} and 
\eqref{eq:CPM_Minkowskii_Aadust}, respectively. 
These terms contribute to reduce the violations of the 
$\mathcal{M}$-constraint.
This is the main consequence of the two adjusted BSSN formulations.

Second, we also find that the violations of the $\mathcal{A}$-constraint 
and $\mathcal{S}$-constraint are larger than those in 
Fig.\ref{fig:ConstraintViolations_GaugeWave}.
From constraint propagation equations \eqref{eq:CPA_Minkowskii} and 
\eqref{eq:CPA}, the violation of the $\mathcal{A}$-constraint is triggered 
by the $\mathcal{M}$- and $\mathcal{A}$-constraints.
The increase in the violations of the $\mathcal{A}$-constraint is caused by
the term $2\lambda_{\widetilde{A}}\delta^{i j}(\partial_i \mathcal{M}_j)$.
Similarly, in \eqref{eq:CPS_Minkowskii} and \eqref{eq:CPS}, the violation of
the $\mathcal{S}$-constraint is triggered by only the 
$\mathcal{A}$-constraint since the magnitude of $\lambda_{\widetilde
{\gamma}}$ is negligible.
Therefore, the increase in the violation of the $\mathcal{S}$-constraint is
due to the violation of the $\mathcal{A}$-constraint.

\begin{figure}[t]
  \begin{center}
    \includegraphics[keepaspectratio=true,width=85mm]{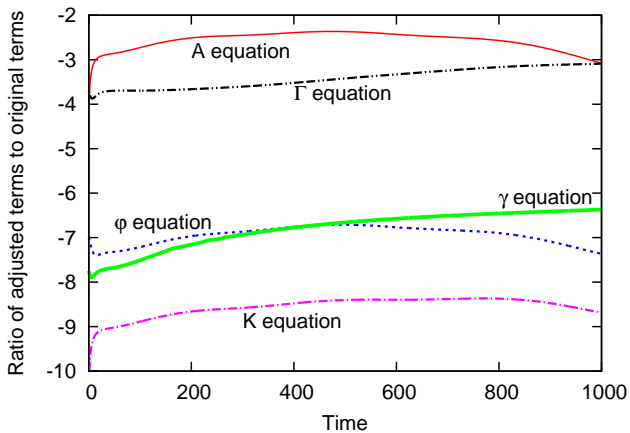}
  \end{center}
  \caption{
    L2 norm of the ratio (adjusted terms)/(original terms) of each evolution
    equation of  the $C^2$-adjusted BSSN formulation,
    \eqref{eq:phi_c2adjusted_BSSN}-\eqref{eq:CGamma_c2adjusted_BSSN}, in 
    the gauge-wave test.
    We see that the largest ratio is the evolution equation of 
    $\widetilde{A}_{i j}$.
    The corrections to $\varphi$, $K$, and $\widetilde{\gamma}_{i j}$ 
    evolution equations are reasonably small.
\label{fig:DiffereneceGauge}}
\end{figure}
From \eqref{eq:appendix_BSSN_DeltaPhi} and 
\eqref{eq:appendix_BSSN_DeltaGamma}, it can be seen that the adjusted terms 
of the evolution equations of $\varphi$ and $\widetilde{\gamma}_{i j}$ 
include second-order derivative terms of the $\mathcal{H}$-constraint.
This means that these evolution equations include fourth-order derivative 
terms of the dynamical variables.
In order to investigate the magnitudes of the adjusted terms, we show in 
Fig.\ref{fig:DiffereneceGauge} the ratio of the adjusted terms to that of 
the original terms in each evolution equation.
We see that the magnitudes of the adjusted terms of $\varphi$ and 
$\widetilde{\gamma}_{i j}$ are reasonably small.

In the simulations with the $C^2$-adjusted BSSN formulation, the largest 
violation is the $\mathcal{S}$-constraint.
The $\mathcal{S}$-constraint depends only on the dynamical variables
$\widetilde{\gamma}_{i j}$, so that there is no other choice than setting 
$\lambda_{\widetilde{\gamma}}$ for controlling $\mathcal{S}$-constraint, as 
can be seen from \eqref{eq:CPS_Minkowskii}.
However, we must set $\lambda_{\widetilde{\gamma}}$ to a value as small as
possible since the adjusted term of $\widetilde{\gamma}_{i j}$ includes 
higher derivatives of $\widetilde{\gamma}_{i j}$.
Therefore, it is hard to control the $\mathcal{S}$-constraint, and we 
have not yet found an appropriate set of parameters.
This will remain as a future problem of this $C^2$-adjusted BSSN system.

We also investigated the sensitivity of the parameters in the 
$C^2$-adjusted BSSN evolutions.
We compared evolutions with setting only one of the parameters, 
$(\lambda_{\varphi}, \lambda_K, \lambda_{\widetilde{\gamma}}, 
\lambda_{\widetilde{A}}, \lambda_{\widetilde{\Gamma}})$, nonzero.
Since the key of the damping of the violation of constraints is the 
$\mathcal{M}$-constraint, and $(\lambda_K, \lambda_{\widetilde{A}})$ 
controls the violation of $\mathcal{M}$-constraint directly by 
\eqref{eq:CPM_Minkowskii}, we mention here only the dependence on 
$\lambda_K$ and $\lambda_{\widetilde{A}}$.
We found that constraint-damping feature changes sensitively by both 
$\lambda_K$ and $\lambda_{\widetilde{A}}$, among them setting 
$\lambda_{\widetilde{A}}$ is important to control the 
$\mathcal{M}$-constraint violation.
We see the best controlled evolution with 
$\lambda_{\widetilde{A}}=10^{-3}$, than $10^{-2}$ and $10^{-4}$.

\subsubsection{Contribution of Algebraic Constraints \\in Definition of 
$C^2$}
\label{C2DefinitionGauge}

\begin{figure}[t]
  \begin{center}
    \includegraphics[keepaspectratio=true,width=85mm]{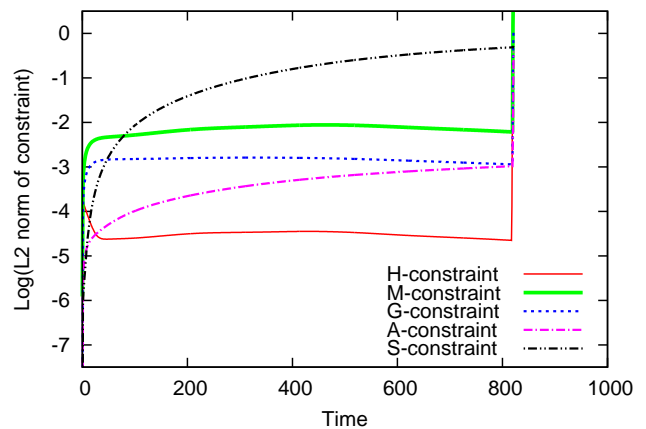}
  \end{center}
  \caption{
    Difference with the definition of $C^2$, \eqref{eq:C2definition}, in 
    the damping of each constraint violation with $c_G=c_A=c_S=0$.
    The parameters $\lambda_{\varphi}$, $\lambda_{K}$, 
    $\lambda_{\widetilde{\gamma}}$, $\lambda_{\widetilde{A}}$, and 
    $\lambda_{\widetilde{\Gamma}}$ are the same as those in 
    Fig.\ref{fig:C2_GaugeWave}.
    The simulation stops since the violations of the constraints sudden 
    increase at $t=800$.
    \label{fig:C2definition_GaugeWave}}
\end{figure}
In Sec.\ref{c2adjustedBSSN}, we defined $C^2$, \eqref{eq:C2definition}, 
including the algebraic constraints.
We check this validity by turning off the algebraic constraints in 
\eqref{eq:C2definition}.
The result is shown in Fig.\ref{fig:C2definition_GaugeWave}, where we see
the simulation stops at $t=800$ due to a sudden increase in the 
violation of the constraints.
This confirms that the algebraic constraints play an important role of 
damping of the violations of constraints.
We also tested with other combinations of Boolean parameters $(c_G, c_A, 
c_S)$, and confirmed that the best controlled evolution is realized when 
$c_G=c_A=c_S=1$.



\subsection{Gowdy-wave Testbed}
\label{GOWDY_TEST}

\subsubsection{Metric and Parameters}
\label{numericalImplimentationGowdy}

The metric of the polarized Gowdy wave is given by
\begin{align}
  d s^2=t^{-1/2} e ^{\lambda/2}(-d t^2+d x^2)+t( e ^P d y^2+ e ^{-P}dz^2),
\end{align}
where $P$ and $\lambda$ are functions of $x$ and $t$.
The forward direction of the time coordinate $t$ corresponds to the 
expanding universe, and $t=0$ corresponds to the cosmological singularity.

For simple forms of the solutions, $P$ and $\lambda$ are given by
\begin{align}
  P &= J_0(2\pi t)\cos(2\pi x),\\
  \lambda 
  &= -2\pi t J_0(2\pi t)J_1(2\pi t)\cos^2(2\pi x)+2\pi^2t^2[J_0^2(2\pi t)
    \nonumber\\
    &\quad
    +J_1^2(2\pi t)]
  -(1/2)\{
  (2\pi)^2[J_0^2(2\pi)+J_1^2(2\pi)]
  \nonumber\\
  &\quad
  -2\pi J_0(2\pi)J_1(2\pi)
  \},
\end{align}
where $J_n$ is the Bessel function.

Following \cite{Alcubierre_etc04}, a new time coordinate $\tau$, which
satisfies harmonic slicing, is obtained by the coordinate transformation 
\begin{align}
  t(\tau) = k  e ^{c\tau},
\end{align}
where $k$ and $c$ are arbitrary constants.
We also follow \cite{Alcubierre_etc04} by setting $k$, $c$, and the initial 
time $t_0 $ as
\begin{align}
  k &\sim 9.67076981276405,\quad
  c \sim 0.002119511921460,\\
  t_0 &=9.87532058290982,
\end{align}
so that the lapse function in the new time coordinate is unity and $t=\tau$ 
at the initial time.

We also use the following parameters specified in \cite{Alcubierre_etc04}.
\begin{itemize}
\item Simulation domain: $x\in[-0.5,0.5], y=z=0$.
  \item Grid: $x_n = -0.5+(n-(1/2))dx$, $n=1,\cdots,100$, where $d x=1/100$.
  \item Time step: $d t=0.25d x$.
  \item Boundary conditions: Periodic boundary condition in $x$-direction 
    and planar symmetry in $y$- and $z$-directions.
  \item Gauge conditions:  $\partial_t\alpha = -\alpha^2 K$, $\beta^i=0$.
  \item Scheme: second-order iterative Crank-Nicolson.
\end{itemize}


\subsubsection{Constraint Violations and Their Dampings}
\label{ConstraintviolationsGowdy}

\begin{figure}[t]
  \begin{center}
    \includegraphics[keepaspectratio=true,width=85mm]{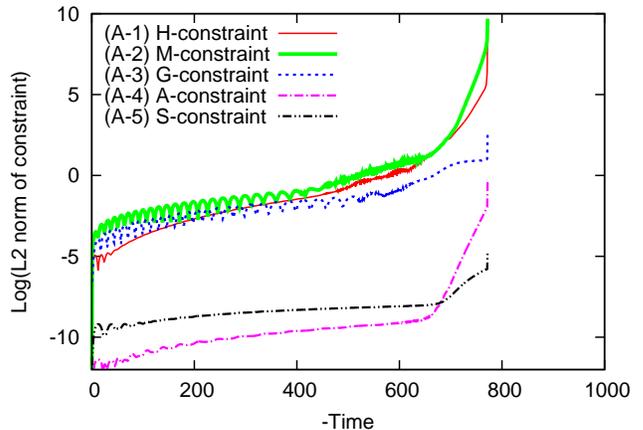}
  \end{center}
  \caption{
    L2 norm of each constraint equation in the polarized Gowdy wave 
    evolution using the standard BSSN formulation.
    The vertical axis is the logarithm of the L2 norm of the constraint and 
    the horizontal axis is backward time.
    \label{fig:ConstraintViolationsGowdy}}
\end{figure}
We begin showing the case of the standard BSSN formulation, 
\eqref{eq:phi_standardBSSNEvolutionEquations}-\eqref{eq:CGamma_standardBSSNEvolutionEquations}.
Figure \ref{fig:ConstraintViolationsGowdy} shows the L2 norm of the
violations of the constraints as a function of backward time $(-t)$.
We see that the violation of the $\mathcal{M}$-constraint is the largest at
all times and that all the violations of constraints increase monotonically 
with time. 
[Comparing with the result in \cite{KS08}, our code shows that the 
$\mathcal{H}$-constraint (A-1) remains at the same level but the 
$\mathcal{M}$-constraint (A-2) is smaller.]

\begin{figure}[t]
  \begin{center}
    \includegraphics[keepaspectratio=true,width=85mm]{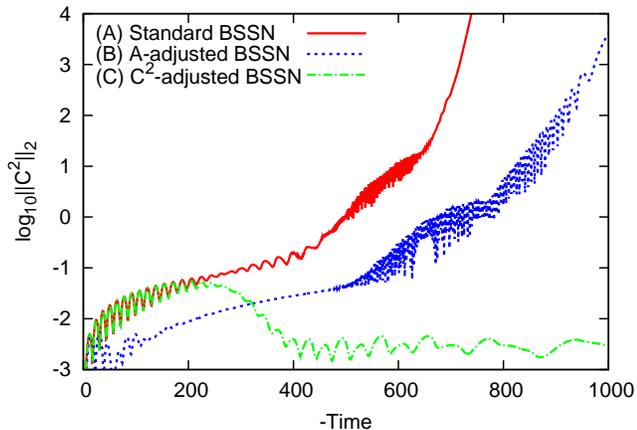}
  \end{center}
  \caption{
    L2 norm of the constraints, $C^2$, of the polarized Gowdy wave tests for
    the standard BSSN and two adjusted formulations.
    The vertical axis is the logarithm of the L2 norm of $C^2$ and the
    horizontal axis is backward time.
    The solid line (A) is the standard BSSN formulation, the dotted line (B)
    is the $\widetilde{A}$-adjusted BSSN formulation with $\kappa_A = 
    -10^{-0.2}$, and the dot-dashed line (C) is the $C^2$-adjusted BSSN 
    formulation with $\lambda_{\varphi} = -10^{-10}$, $\lambda_{K} = 
    -10^{-4.6}$, $\lambda_{\widetilde{\gamma}} = -10^{-11}$, 
    $\lambda_{\widetilde{A}} = -10^{-1.2}$, and $\lambda_{\widetilde{\Gamma}} = 
    -10^{-14.3}$.
    Note that the signatures of $\kappa_A$ and $\lambda$s are negative since
    the simulations evolve backward.
    We see that lines (A) and (C) are identical until $t=-200$.
    Line (C) then decreases and maintains its magnitude under $O(10^{-2})$ 
    after $t=-400$.
    We confirm this behavior until $t=-1500$.
    \label{fig:DampingViolations_Gowdy}}
\end{figure}
Similar to the gauge-wave test, we compare the violations of $C^2$ for three
types of BSSNs in Fig.\ref{fig:DampingViolations_Gowdy}. 
In the case of the $\widetilde{A}$-adjusted BSSN formulation, the violation 
of the constraints increases if we set $|\kappa_A|$ larger than $10^{-0.2}$.
In the case of the $C^2$-adjusted BSSN formulation, it increases if we set 
$|\lambda_{\widetilde{A}}|$ larger than $10^{-1.2}$.
Note that the signatures of the above $\kappa_A$ and $\lambda$s are 
negative, contrary to the predictions in \cite{YS02} and 
Sec.\ref{ApplicationEinsteinEq}, respectively.
This is because these simulations are performed with backward time.

\begin{figure}[t]
  \begin{center}
    \includegraphics[keepaspectratio=true,width=85mm]{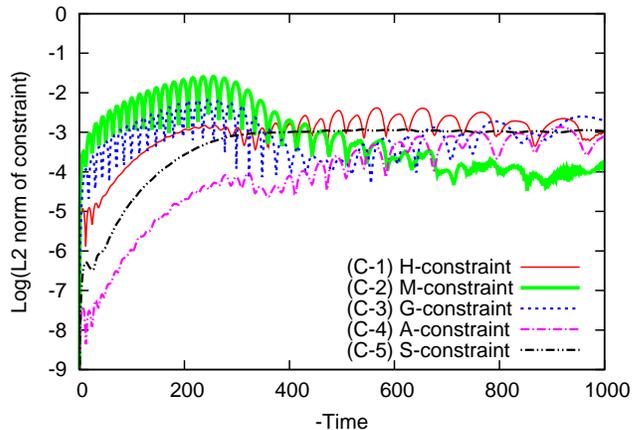}
  \end{center}
  \caption{
    The same with Fig.\ref{fig:ConstraintViolationsGowdy} but for the 
    $C^2$-adjusted BSSN formulation.
    The parameters, ($\lambda_{\varphi}$, $\lambda_{K}$, 
    $\lambda_{\widetilde{\gamma}}$, $\lambda_{\widetilde{A}}$,
    $\lambda_{\widetilde{\Gamma}}$), are the same with those for (C) in 
    Fig.\ref{fig:DampingViolations_Gowdy}.
    We see that the violation of the $\mathcal{M}$-constraint decreases and
    becomes the lowest after $t=-700$.
    \label{fig:C2Constraints}}
\end{figure}
As shown in Fig.\ref{fig:DampingViolations_Gowdy}, the violations of $C^2$
for the standard BSSN formulation and the $\widetilde{A}$-adjusted BSSN 
formulation increase monotonically with time, while that for the 
$C^2$-adjusted BSSN formulation decreases after $t=-200$.
To investigate the reason of this rapid decay after $t=-200$, we plot each 
constraint violation in Fig.\ref{fig:C2Constraints}. 
We see that the violations of the $\mathcal{A}$-constraint and 
$\mathcal{S}$-constraint increase with negative time, in contrast to the
standard BSSN formulation, and those of the $\mathcal{M}$-constraint and 
$\mathcal{G}$-constraint decrease after $t=-200$.
The propagation equation of the $\mathcal{M}$-constraint,
\eqref{eq:CPM_Minkowskii}, includes the term $-2c_A\lambda_{\widetilde{A}}
\partial_a \mathcal{A}$, which contributes to constraint damping.
Similarly, the propagation equation of the $\mathcal{G}$-constraint, 
\eqref{eq:CPG_Minkowskii}, includes $\delta^{a b}\{(1/2)
\lambda_{\widetilde{\gamma}}\partial_b\Delta + 2\lambda_{\widetilde{\Gamma}}
\partial_b\}\mathcal{H} -c_S\lambda_{\widetilde{\gamma}}\delta^{a b}
\partial_b \mathcal{S}$; the decay of the violations of the 
$\mathcal{G}$-constraint is caused by these terms.
Therefore, these terms are considered to become significant of approximately
$t=-200$ when the violations of the $\mathcal{A}$, $\mathcal{H}$, and 
$\mathcal{S}$-constraints become a certain order of magnitude.

\begin{figure}[t]
  \begin{center}
    \includegraphics[keepaspectratio=true,width=85mm]{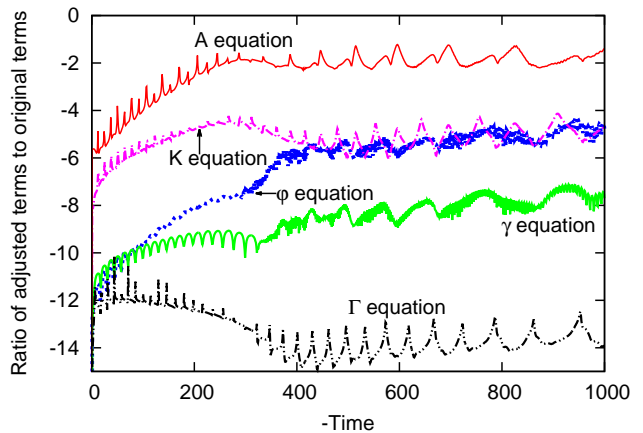}
  \end{center}
  \caption{
    L2 norm of the ratio (adjusted terms)/(original terms) of each evolution
    equation for the $C^2$-adjusted BSSN formulation, 
    \eqref{eq:phi_c2adjusted_BSSN}-\eqref{eq:CGamma_c2adjusted_BSSN}.
    We see that the largest ratio is that for the evolution of 
    $\widetilde{A}_{i j}$.
    The corrections to the $\widetilde{\gamma}_{i j}$ and 
    $\widetilde{\Gamma}^i$ evolution equations are reasonably small.
    \label{fig:MagnitudeTermsGowdy}}
\end{figure}
In contrast to the gauge-wave testbed (Fig.\ref{fig:DiffereneceGauge}), we 
prepared Fig.\ref{fig:MagnitudeTermsGowdy}, which shows the magnitudes of 
the ratio of the adjusted terms to the original terms.
Since the magnitudes of the adjusted terms of $\varphi$ and
$\widetilde{\gamma}_{i j}$ can be disregarded, the effect of the reduction 
of the adjusted terms of $\varphi$ and $\widetilde{\gamma}_{i j}$ is 
negligible.
Therefore, the $C^2$-adjusted BSSN evolution in the Gowdy wave can be 
regarded as maintaining its original hyperbolicity.




We repeated the parameter-dependency survey of $(\lambda_{\varphi},
\lambda_{K}, \lambda_{\widetilde{\gamma}},\lambda_{\widetilde{A}}, 
\lambda_{\widetilde{\Gamma}})$ for this spacetime evolution.
Similar to Sec.\ref{ConstraintviolationsGauge}, we found that 
constraint-damping feature is sensitive to both $\lambda_K$ and 
$\lambda_{\widetilde{A}}$, of which $\lambda_{\widetilde{A}}$ works
effectively than $\lambda_K$.
We see the most controlled evolution when $\lambda_{\widetilde{A}}
=10^{-1}$, than that of $\lambda_{\widetilde{A}}=10^0$ or $\lambda_
{\widetilde{A}}=10^{-2}$.


\subsubsection{Contribution of Algebraic Constraints \\in Definition of
$C^2$}
\label{C2DefinitionGowdy}

\begin{figure}[t]
  \begin{center}
    \includegraphics[keepaspectratio=true,width=85mm]{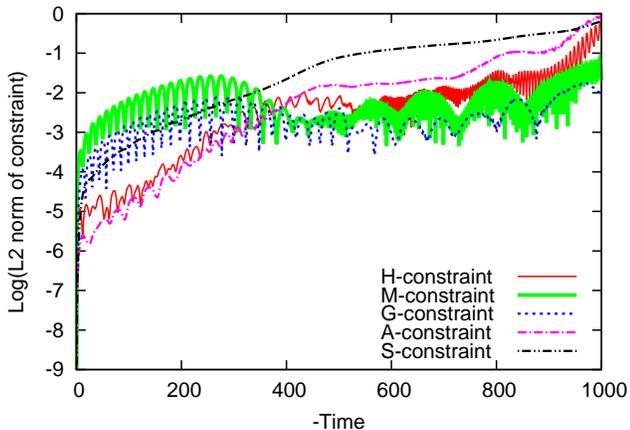}
  \end{center}
  \caption{
    Difference with the definition of $C^2$ with $c_G=c_A=c_S=0$.
    The coefficient parameters, $\lambda_{\varphi}$, $\lambda_{K}$, 
    $\lambda_{\widetilde{\gamma}}$, $\lambda_{\widetilde{A}}$ and 
    $\lambda_{\widetilde{\Gamma}}$, are all the same as those for (C) in
    Fig.\ref{fig:DampingViolations_Gowdy}.
    In comparison with Fig.\ref{fig:C2Constraints}, all the violations of
    the constraints are larger.
    \label{fig:C2Parameter_GowdyWave2}}
\end{figure}
In Sec.\ref{c2adjustedBSSN}, we investigated the effect of the definition of
$C^2$. 
Similar to the gauge-wave tests in the previous subsection, we show 
the effect of constraint damping caused by the algebraic constraints.
In Fig.\ref{fig:C2Parameter_GowdyWave2}, we plot the violations of all the 
constraint with $c_G=c_A=c_S=0$.
We see that all the violations of the constraints are larger than those in 
Fig.\ref{fig:C2Constraints}.
This result is consistent with the discussion in Sec.\ref{c2adjustedBSSN}.




\section{Summary and Discussion}
\label{Summary}

To obtain an evolution system robust against the violation of constraints, 
we derived a new set of adjusted BSSN equations applying the idea proposed 
by Fiske \cite{Fiske04} which we call a ``$C^2$-adjusted system.''
That is, we added the functional derivatives of the norm of the constraints,
$C^2$, to the evolution equations
[\eqref{eq:phi_c2adjusted_BSSN}-\eqref{eq:CGamma_c2adjusted_BSSN}].
We performed numerical tests in the gauge-wave and Gowdy wave spacetimes
and confirmed that the violations of constraints decrease as expected, 
and that longer and accurate simulation than that of the standard BSSN 
evolution is available.

The construction of the $C^2$-adjusted system is straightforward.
However, in BSSN, there are two kinetic constraints and three additional 
algebraic constraints compared to the ADM system; thus, the definition of 
$C^2$ is a matter of concern.
By analyzing constraint propagation equations, we concluded that $C^2$ 
should include all the constraints.
This was also confirmed by numerical tests.
The importance of such algebraic constraints suggests the similar
treatment when we apply this idea to other formulations of the Einstein
equation.

To evaluate the reduction of the violations of the constraints, we also 
compared evolutions with the 
$\widetilde{A}$-adjusted BSSN formulation proposed in \cite{YS02}.
We concluded that the $C^2$-adjusted BSSN formulation exhibits superior 
constraint damping to both the standard and $\widetilde{A}$-adjusted BSSN 
formulations.
In particular, the lifetimes of the simulations of the $C^2$-adjusted BSSN
formulation  in the gauge-wave and Gowdy wave testbeds are as ten-times and 
twice as longer than those of the standard BSSN formulation, respectively.

So far, many trials have been reported to improve BSSN formulation
(e.g. \cite{YS02, BH10}).
Recently, for example, a conformal-traceless Z4 formulation was proposed 
with its test demonstrations \cite{ABBRP11}.
Among them, Fig.1 of \cite{ABBRP11} can be compared with our 
Fig.\ref{fig:ConstraintViolationsC2_GaugeWave} [(B-1) and (C-1)] as the
same gauge-wave test.
The violation of $\mathcal{H}$-constraint in $C^2$-adjusted evolution 
looks smaller than that of new Z4 evolution, but regarding the blow-up 
time of simulations, new Z4 system has advantage.

Fiske reported the applications of the idea of $C^2$-adjustment to 
``linearized'' ADM and BSSN formulations in his dissertation 
\cite{Fiske_Phd}.  
(As he mentioned, his BSSN is not derived from the standard BSSN equations
but from a linearized ADM using a new variable, $\Gamma$.
His set of BSSN equations also does not include the $\mathcal{A}$- and 
$\mathcal{S}$-constraints in our notation.).  
He observed damping of the constraint violation of five orders of magnitude
and the equivalent solution errors in his numerical evolution tests. 
Our studies show that the full BSSN set of 
equations with fully adjusted terms also produces the desired 
constraint-damping results (Fig.\ref{fig:C2_GaugeWave} and 
Fig.\ref{fig:DampingViolations_Gowdy}), 
although apparent improvements are at fewer orders of magnitude.

When applied this idea to the ADM system \cite{TYS11}, we found  that the 
adjustment to the $K_{ij}$-evolution equation is essential.  
In the present study, we found that the adjustment to the 
$\widetilde{A}_{ij}$-evolution equation is essential for controlling the 
constraints.  
In both cases, the associated adjustment parameters 
(Lagrangian multipliers), $\lambda_{\widetilde{A}}$ in this study, are 
sensitive and require fine-tuning.
In future, automatic controlling system such that 
monitoring the order of constraint violations and maintaining 
them by tuning the parameters automatically would be helpful. 
Applications of control theory in this direction are being investigated.

The correction terms of the $C^2$-adjusted system include higher-order 
derivatives and are not quasi-linear; thus, little is known mathematically
about such systems. 
These additional terms might effectively act as artificial viscosity terms 
in fluid simulations, but might also enhance the violation of errors. 
To investigate this direction further, the next step is to apply the idea to
a system in which constraints do not include second-order derivatives of 
dynamical variables.  We are working on the Kidder-Scheel-Teukolsky 
formulation \cite{KST01} as an example of such a system, which we will 
report in the near future.


\begin{acknowledgments}
This work was partially supported by Grant-in-Aid for
Scientific Research Fund of Japan Society of the Promotion of Science
No. 22540293 (HS).
Numerical computations were carried out on an Altix 3700 BX2 supercomputer 
at YITP in Kyoto University and on the RIKEN Integrated Cluster of Clusters
(RICC).
\end{acknowledgments}

\appendix

\section{Additional $C^2$-adjusted Terms}
\label{Appendix_adjustTermsC2}

The adjusted terms $\delta C^2/\delta \varphi$, $\delta C^2/\delta K$,
$\delta C^2/\delta \widetilde{\gamma}_{m n}$, 
$\delta C^2/\delta \widetilde{A}_{m n}$, and 
$\delta C^2/\delta \widetilde{\Gamma}^a$ in
\eqref{eq:phi_c2adjusted_BSSN}-\eqref{eq:CGamma_c2adjusted_BSSN} are written
as follows:
\begin{widetext}
\begin{align}
\frac{\delta C^2}{\delta \varphi}
&=
2\bar{H}_1\mathcal{H}
-2(\partial_a\bar{H}_2^a)\mathcal{H}
-2\bar{H}_2^a\partial_a\mathcal{H}
+2(\partial_a\partial_b\bar{H}_3^{a b})\mathcal{H}
+2(\partial_a\bar{H}_3^{a b})\partial_b\mathcal{H}
+2(\partial_b\bar{H}_3^{a b})\partial_a\mathcal{H}
+2\bar{H}_3^{a b}\partial_a\partial_b\mathcal{H}
\nonumber\\
&\quad
-2(\partial_a\bar{M}_{1 i}{}^a) e ^{-4\varphi}
\widetilde{\gamma}^{i j}\mathcal{M} _j
+8\bar{M}_{1 i}{}^a e ^{-4\varphi}(\partial_a\varphi)
\widetilde{\gamma}^{i j}\mathcal{M} _j
-2\bar{M}_{1 i}{}^a e ^{-4\varphi}
(\partial_a\widetilde{\gamma}^{i j})\mathcal{M} _j
-2\bar{M}_{1 i}{}^a e ^{-4\varphi}
\widetilde{\gamma}^{i j}\partial_a\mathcal{M} _j
\nonumber\\
&\quad
-4\widetilde{\gamma}^{i j} e ^{-4\varphi}
\mathcal{M} _i\mathcal{M} _j
+4c_G  e ^{4\varphi}\widetilde{\gamma}_{i j}\mathcal{G}^i
\mathcal{G}^j,
\label{eq:appendix_BSSN_DeltaPhi}
\\
\frac{\delta C^2}{\delta K}
&=
2\bar{H}_4\mathcal{H}
-2(\partial_\ell\bar{M}_{2 i}{}^\ell) e ^{-4\varphi}\widetilde{\gamma}^{i j}
\mathcal{M} _j
+8\bar{M}_{2 i}{}^\ell e ^{-4\varphi}(\partial_\ell \varphi)
\widetilde{\gamma}^{i j}\mathcal{M} _j
-2\bar{M}_{2 i}{}^\ell e ^{-4\varphi}(\partial_\ell \widetilde{\gamma}^{i j})
\mathcal{M} _j
-2\bar{M}_{2 i}{}^\ell e ^{-4\varphi}\widetilde{\gamma}^{i j}
\partial_\ell \mathcal{M} _j,
\label{eq:appendix_BSSN_DeltaK}\\
\frac{\delta C^2}{\delta \widetilde{\gamma}_{m n}}
&=
2\bar{H}_5^{m n}\mathcal{H}
-2(\partial_i\bar{H}_6^{i m n})\mathcal{H}
-2\bar{H}_6^{i m n}\partial_i\mathcal{H}
+2(\partial_i\partial_j\bar{H}_7^{i j m n})\mathcal{H}
+2(\partial_i\bar{H}_7^{i j m n})\partial_j\mathcal{H}
+2(\partial_j\bar{H}_7^{i j m n})\partial_i\mathcal{H}
\nonumber\\
&\quad
+2\bar{H}_7^{i j m n}\partial_i\partial_j\mathcal{H}
+2\bar{M}_{3 i}{}^{m n} e ^{-4\varphi}\widetilde{\gamma}^{i j}
\mathcal{M} _j
-2(\partial_c\bar{M}_{4 i}{}^{c m n}) e ^{-4 \varphi}
\widetilde{\gamma}^{i j}\mathcal{M} _j
+8\bar{M}_{4 i}{}^{c m n} e ^{-4 \varphi}(\partial_c\varphi)
\widetilde{\gamma}^{i j}\mathcal{M} _j
\nonumber\\
&\quad
-2\bar{M}_{4 i}{}^{c m n} e ^{-4 \varphi}
(\partial_c\widetilde{\gamma}^{i j})\mathcal{M} _j
-2\bar{M}_{4 i}{}^{c m n} e ^{-4 \varphi}
\widetilde{\gamma}^{i j}\partial_c\mathcal{M} _j
- e ^{-4\varphi}\widetilde{\gamma}^{i m}\widetilde{\gamma}^{j n}
\mathcal{M} _i\mathcal{M} _j
+2c_G G_1^{i m n} e ^{4\varphi}\widetilde{\gamma}_{i j}
\mathcal{G}^j
\nonumber\\
&\quad
-2c_G (\partial_\ell G_2^{i m n \ell}) e ^{4\varphi}
\widetilde{\gamma}_{i j}\mathcal{G}^j
-8c_G G_2^{i m n \ell} e ^{4\varphi}(\partial_\ell \varphi)
\widetilde{\gamma}_{i j}\mathcal{G}^j
-2c_G G_2^{i m n \ell} e ^{4\varphi}
(\partial_\ell \widetilde{\gamma}_{i j})\mathcal{G}^j
-2c_G G_2^{i m n \ell} e ^{4\varphi}
\widetilde{\gamma}_{i j}\partial_\ell \mathcal{G}^j
\nonumber\\
&\quad
+c_G  e ^{4\varphi}\mathcal{G}^m\mathcal{G}^n
+2c_A A_1^{m n}\mathcal{A}
+2c_SS_1^{m n}\mathcal{S},
\label{eq:appendix_BSSN_DeltaGamma}\\
\frac{\delta C^2}{\delta \widetilde{A}_{m n}}
&=2\bar{H}_8^{m n}\mathcal{H}
+2 e ^{-4\varphi}\widetilde{\gamma}^{i j}\bar{M}_{5 i}{}^{m n}
\mathcal{M} _j
-2(\partial_c\bar{M}_{6 i}{}^{c m n}) e ^{-4\varphi}
\widetilde{\gamma}^{i j}\mathcal{M} _j
+8\bar{M}_{6 i}{}^{c m n} e ^{-4\varphi}(\partial_c\varphi)
\widetilde{\gamma}^{i j}\mathcal{M} _j
\nonumber\\
&\quad
-2\bar{M}_{6 i}{}^{c m n} e ^{-4\varphi}
(\partial_c\widetilde{\gamma}^{i j})\mathcal{M} _j
-2\bar{M}_{6 i}{}^{c m n} e ^{-4\varphi}
\widetilde{\gamma}^{i j}\partial_c\mathcal{M} _j
+2c_A A_2^{m n}\mathcal{A},
\label{eq:appendix_BSSN_DeltaA}\\
\frac{\delta C^2}{\delta \widetilde{\Gamma}^a}
&=
2\bar{H}_{9 a}\mathcal{H}
-2(\partial_b\bar{H}_{10a}^b)\mathcal{H}
-2\bar{H}_{10a}^b\partial_b\mathcal{H}
+2c_G G_{3 a}^i e ^{4\varphi}\widetilde{\gamma}_{i
j}\mathcal{G}^j,
\label{eq:appendix_BSSN_DeltaCGamma}
\end{align}
\end{widetext}
where
\begin{align}
\bar{H}_1
&= -4 e ^{-4\varphi}\widetilde{R}
+32 e ^{-4\varphi}\{\widetilde{D}^{i}\widetilde{D}_{i} \varphi
+(\widetilde{D}_{i} \varphi)(\widetilde{D}^{i} \varphi)\},
\label{eq:appendix_BSSN_H1}
\end{align}
\begin{align}
\bar{H}_2^a
&= 8 e ^{-4\varphi}(
\widetilde{\gamma}^{i j}\widetilde{\Gamma}^a{}_{i j}
-2\widetilde{D}^{a}\varphi),
\label{eq:appendix_BSSN_H2}
\end{align}
\begin{align}
\bar{H}_3^{a b}
&= -8 e ^{-4\varphi}\widetilde{\gamma}^{a b},
\label{eq:appendix_BSSN_H3}
\end{align}
\begin{align}
\bar{H}_4&=
(4/3)K-(2/3)\widetilde{\gamma}^{i j}\widetilde{A}_{i j},
\label{eq:appendix_BSSN_H4}
\end{align}
\begin{align}
\bar{H}_5^{m n}
&=
- e ^{-4\varphi}\widetilde{R}^{m n}
+ e ^{-4\varphi}(\partial_j\widetilde{\Gamma}^{(m})\widetilde{\gamma}^{n) j}
\nonumber\\
&\quad
-2 e ^{-4\varphi}\widetilde{\Gamma}^{k m}{}_j\widetilde{\Gamma}^{jn}{}_{k}
-2 e ^{-4\varphi}\widetilde{\Gamma}^{i \ell(m}\widetilde{\Gamma}^{n)}{}_{\ell i}
\nonumber\\
&\quad
- e ^{-4\varphi}\widetilde{\Gamma}^{a m i}\widetilde{\Gamma}_{a i}{}^n 
- e ^{-4\varphi}\widetilde{\Gamma}^{m i \ell}\widetilde{\Gamma}^{n}{}_{\ell i}
\nonumber\\
&\quad
+(1/2) e ^{-4\varphi}\widetilde{\gamma}_{ij,a\ell}\widetilde{\gamma}^{i j}
\widetilde{\gamma}^{a m}\widetilde{\gamma}^{\ell n}
+8 e ^{-4\varphi}\widetilde{D}^m\widetilde{D}^n\varphi
\nonumber\\
&\quad
-8 e ^{-4\varphi}(\widetilde{D}^{(m}\varphi)\widetilde{\Gamma}^{n)}{}_{i j}
\widetilde{\gamma}^{i j}
+8 e ^{-4\varphi}(\widetilde{D}^m\varphi)(\widetilde{D}^n\varphi)
\nonumber\\
&\quad
+2\widetilde{A}^{m b}\widetilde{A}^n{}_b
+(2/3)\widetilde{A}^{m n}K,
\label{eq:appendix_BSSN_H5}
\end{align}
\begin{align}
\bar{H}_6^{\ell m n}
&= e ^{-4\varphi}
\{\widetilde{\Gamma}^{\ell m n}
+2\widetilde{\Gamma}^{(n m)\ell}
+(1/2)\Gamma^\ell\widetilde{\gamma}^{m n}
\nonumber\\
&\quad
+8\widetilde{\gamma}^{
\ell (m}(\widetilde{D}^{n)}\varphi)
-4\widetilde{\gamma}^{m
n}\widetilde{D}^\ell \varphi\},
\label{eq:appendix_BSSN_H6}
\end{align}
\begin{align}
\bar{H}_7^{i j m n}
&= -(1/2) e ^{-4\varphi}
\widetilde{\gamma}^{ m n}\widetilde{\gamma}^{i j},
\label{eq:appendix_BSSN_H7}
\end{align}
\begin{align}
\bar{H}_8^{m n}
&=
-2\widetilde{A}^{m n}
-(2/3)\widetilde{\gamma}^{m n}K,
\label{eq:appendix_BSSN_H8}
\end{align}
\begin{align}
\bar{H}_9{}_a&=
(1/2) e ^{-4\varphi}\widetilde{\gamma}^{i j}\widetilde{\gamma}_{i
j,a},
\label{eq:appendix_BSSN_H9}
\end{align}
\begin{align}
 \bar{H}_{10}^b{}_a
&= e ^{-4\varphi}\delta^b{}_a,
\label{eq:appendix_BSSN_H10}
\end{align}
\begin{align}
\bar{M}_{1 i}{}^a
&=6\widetilde{A}^a{}_i-2\widetilde{A}_{m n}\widetilde{\gamma}^{m n}
\delta^a{}_i,
\label{eq:appendix_BSSN_M1}
\end{align}
\begin{align}
\bar{M}_{2 i}{}^j
&=-(2/3)\delta^j{}_i,
\label{eq:appendix_BSSN_M2}
\end{align}
\begin{align}
\bar{M}_{3 i}{}^{m n}
&=
-6(\widetilde{D}^{(m}\varphi)\widetilde{A}^{n)}{}_i
+2(\widetilde{D}_i \varphi)\widetilde{A}^{m n}
-\widetilde{D}^{(m}\widetilde{A}^{n)}{}_{i}
\nonumber\\
&\quad
+\widetilde{A}^{a (n}\widetilde{\Gamma}^{m)}{}_{a i}
+\widetilde{A}_i{}^{(m}\widetilde{\Gamma}^{n)}{}_{j \ell}
\widetilde{\gamma}^{j \ell},
\label{eq:appendix_BSSN_M3}
\end{align}
\begin{align}
\bar{M}_{4 i}{}^{c m n}
&=
-\widetilde{\gamma}^{c (n}\widetilde{A}^{m)}{}_i
+(1/2)\widetilde{\gamma}^{m n}\widetilde{A}^c{}_i
-(1/2)\widetilde{A}^{n m}\delta^c{}_i,
\label{eq:appendix_BSSN_M4}
\end{align}
\begin{align}
\bar{M}_{5 i}{}^{m n}
&=
6(\widetilde{D}^{(m} \varphi)\delta^{n)}{}_i
-2(\widetilde{D}_i \varphi)\widetilde{\gamma}^{m n}
-\delta_i{}^{(m}\widetilde{\Gamma}^{n)}{}_{j \ell}\widetilde{\gamma}^{j \ell}
\nonumber\\
&\quad
+(1/2)\widetilde{\gamma}{}^{m n}{}_{,i},
\label{eq:appendix_BSSN_M5}
\end{align}
\begin{align}
\bar{M}_{6 i}{}^{c m n}
&=
\widetilde{\gamma}^{c (m}\delta^{n)}{}_i,
\label{eq:appendix_BSSN_M6}
\end{align}
\begin{align}
G_1^{i a b}
&=
\widetilde{\Gamma}^{i a b}
+\widetilde{\gamma}^{i (b} \widetilde{\Gamma}^{a)}{}_{m
n}\widetilde{\gamma}^{m n},
\label{eq:appendix_BSSN_G1}
\end{align}
\begin{align}
G_2^{i a b \ell}
&=
-\widetilde{\gamma}^{\ell (b}\widetilde{\gamma}^{a) i}
+(1/2)\widetilde{\gamma}^{a b}\widetilde{\gamma}^{i \ell},
\label{eq:appendix_BSSN_G2}
\end{align}
\begin{align}
G_3^i{}_j
&=
\delta^i{}_j,
\label{eq:appendix_BSSN_G3}
\end{align}
\begin{align}
A_1^{a b}&=
-\widetilde{A}^{a b},
\label{eq:appendix_BSSN_A1}
\end{align}
\begin{align}
A_2^{a b}&=
\widetilde{\gamma}^{a b},
\label{eq:appendix_BSSN_A2}
\end{align}
\begin{align}
S_1^{a b}
&=
(1/2)\varepsilon^{a j k}\varepsilon^{b n \ell}\widetilde{\gamma}_{jn}
\widetilde{\gamma}_{k\ell}.
\label{eq:appendix_BSSN_S1}
\end{align}

\section{Constraint Propagation Equations of Adjusted BSSN Formulations}
\label{Appendix_CP_Minkowskii}

Here we give the constraint propagation equations for the $C^2$-adjusted
BSSN formulation and the $\widetilde{A}$-adjusted BSSN formulation in 
Minkowskii spacetime.
For simplicity, we set  $\lambda_{\widetilde{\gamma} i j m n}
=\lambda_{\widetilde{\gamma}} \delta_{i m}\delta_{j n}$, 
$\lambda_{\widetilde{A} i j m n} =\lambda_{\widetilde{A}} 
\delta_{i m}\delta_{j n}$, and $\lambda_{\widetilde{\Gamma}}^{i j} 
=\lambda_{\widetilde{\Gamma}} \delta^{i j}$.
The constraint propagation equations of the $C^2$-adjusted BSSN formulation
are
\begin{widetext}
\begin{align}
\partial_t\mathcal{H}
&=
[{\rm Original\,\, Terms}]
+\left(-128\lambda_\varphi\Delta^2
-(3/2)\lambda_{\widetilde{\gamma}}\Delta^2
+2\lambda_{\widetilde{\Gamma}}\Delta\right)\mathcal{H}
+c_G
\left(
-(1/2)\lambda_{\widetilde{\gamma}}\Delta\partial_m
-2\lambda_{\widetilde{\Gamma}}\partial_m
\right)\mathcal{G}^m
+3c_S\lambda_{\widetilde{\gamma}}
\Delta \mathcal{S},
\label{eq:CPH_Minkowskii}\\
\partial_t\mathcal{M}_a
&=
[{\rm Original\,\,Terms}]
+\biggl\{
(8/9)\lambda_K\delta^{b c}\partial_a\partial_b
+\lambda_{\widetilde{A}}\Delta\delta_{a}{}^c
+\lambda_{\widetilde{A}}\delta^{b c}\partial_a\partial_b
\biggr\}\mathcal{M}_c
-2c_A \lambda_{\widetilde{A}}\partial_a\mathcal{A},
\label{eq:CPM_Minkowskii}\\
\partial_t\mathcal{G}^a
&=
[{\rm Original\,\,Terms}]
+\delta^{a b}\left(
(1/2)\lambda_{\widetilde{\gamma}}
\partial_b\Delta
+2\lambda_{\widetilde{\Gamma}}\partial_b
\right)\mathcal{H}
+c_G \left(
\lambda_{\widetilde{\gamma}}
\Delta \delta^a{}_b
+(1/2)\lambda_{\widetilde{\gamma}}\delta^{a c}\partial_c\partial_b
-2\lambda_{\widetilde{\Gamma}}\delta^{a}{}_b
\right)\mathcal{G}^b
-c_S\lambda_{\widetilde{\gamma}} \delta^{a b}\partial_b
\mathcal{S},
\label{eq:CPG_Minkowskii}\\
\partial_t\mathcal{A}
&=
[{\rm Original\,\, Terms}]
+2\lambda_{\widetilde{A}}\delta^{i j}(\partial_i \mathcal{M}_j)
-6c_A \lambda_{\widetilde{A}}\mathcal{A},
\label{eq:CPA_Minkowskii}\\
\partial_t\mathcal{S}
&=
[{\rm Original\,\,Terms}]
+3\lambda_{\widetilde{\gamma}}\Delta \mathcal{H}
+c_G \lambda_{\widetilde{\gamma}}\partial_\ell \mathcal{G}^\ell
-6c_S \lambda_{\widetilde{\gamma}}\mathcal{S},
\label{eq:CPS_Minkowskii}
\end{align}
\end{widetext}
and those of the $\widetilde{A}$-adjusted BSSN formulation are
\begin{align}
\partial_t\mathcal{H}
&=[{\rm Original\,\,Terms}],
\label{eq:CPH_Minkowskii_Aadust}\\
\partial_t\mathcal{M}_i
&=[{\rm Original\,\,Terms}]
+(1/2)\kappa_A \Delta \mathcal{M}_i,
\label{eq:CPM_Minkowskii_Aadust}\\
\partial_t\mathcal{G}^i
&=[{\rm Original\,\,Terms}],
\label{eq:CPG_Minkowskii_Aadust}\\
\partial_t\mathcal{A}
&=[{\rm Original\,\,Terms}]
+\kappa_A \delta^{i j}\partial_i\mathcal{M}_j,
\label{eq:CPA_Minkowskii_Aadust}\\
\partial_t\mathcal{S}
&=[{\rm Original\,\,Terms}],
\label{eq:CPS_Minkowskii_Aadust}
\end{align}
where $\Delta$ is the Laplacian operator in flat space. 
``Original Terms'' refers to the right-hand side of the constraint 
propagation equations for the standard BSSN formulation.
Full expressions for the terms are given in the appendix of \cite{YS02}.

\section{Constraint Propagation Equations of Standard BSSN Formulation 
with $\beta^i=0$}
\label{Appendix_CP_noshit}

The constraint propagation equations for the standard BSSN formulation with
$\beta^i=0$ are as follows (the full expressions are available in the 
appendix of 
\cite{YS02}).
\begin{widetext}
\begin{align}
  \partial_t \mathcal{H}
  &= [(2/3)\alpha K + (2/3)\alpha \mathcal{A}]\mathcal{H}
  +[-4 e^{-4\varphi} \alpha(\alpha_k \varphi) \widetilde{\gamma}^{k j}
    -2e ^{-4\varphi}(\partial_k \alpha)\widetilde{\gamma}^{jk}
  ]\mathcal{M}_j
  \nonumber\\
  &\quad
  +[-2\alpha e^{-4\varphi}\widetilde{A}^k{}_j\partial_k 
    -\alpha e^{-4\varphi}(\partial_j\widetilde{A}_{k\ell})
    \widetilde{\gamma}^{k\ell}
    -e^{-4\varphi}(\partial_j\alpha)\mathcal{A}
  ]\mathcal{G}^j
  \nonumber\\
  &\quad
  +[2\alpha e^{-4\varphi}\widetilde{\gamma}^{-1}
    \widetilde{\gamma}^{\ell k}(\partial_\ell \varphi)\mathcal{A}\partial_k 
    + (1/2)\alpha e^{-4\varphi}\widetilde{\gamma}^{-1}(\partial_\ell
    \mathcal{A})\widetilde{\gamma}^{\ell k}\partial_k
    +(1/2)e^{-4\varphi}\widetilde{\gamma}^{-1}(\partial_\ell \alpha)
    \widetilde{\gamma}^{\ell k}\mathcal{A}\partial_k
  ]\mathcal{S}
  \nonumber\\
  &\quad
  +[(4/9)\alpha K \mathcal{A} 
    - (8/9)\alpha K^2 
    +(4/3)\alpha e^{-4\varphi}(\partial_i \partial_j \varphi)
    \widetilde{\gamma}^{i j}
    +(8/3)\alpha e^{-4\varphi}(\partial_k \varphi)(\partial_\ell
    \widetilde{\gamma}^{\ell k})
    +\alpha e^{-4\varphi}(\partial_j \widetilde{\gamma}^{j k})\partial_k
    \nonumber\\
    &\qquad
    +8\alpha e^{-4\varphi}\widetilde{\gamma}^{j k}(\partial_j\varphi)
    \partial_k
    +\alpha e^{-4\varphi}\widetilde{\gamma}^{j k}\partial_j \partial_k
    +8e^{-4\varphi}(\partial_\ell \alpha)(\partial_k \varphi)
    \widetilde{\gamma}^{\ell k}
    +e^{-4\varphi}(\partial_\ell \alpha)(\partial_k 
    \widetilde{\gamma}^{\ell k})
    +2 e^{-4\varphi}(\partial_\ell \alpha)\widetilde{\gamma}^{\ell k}
    \partial_k
    \nonumber\\
    &\qquad
    +e^{-4\varphi}\widetilde{\gamma}^{\ell k}(\partial_\ell \partial_k
    \alpha)
  ]\mathcal{A},\label{eq:CPH}
\end{align}
\begin{align}
  \partial_t\mathcal{M}_i
  &=[-(1/3)(\partial_i\alpha)+(1/6)\partial_i]\mathcal{H}
  +\alpha K \mathcal{M}_i
  +[\alpha e^{-4\varphi}\widetilde{\gamma}^{k m}(\partial_k
    \varphi)(\partial_j\widetilde{\gamma}_{m i})
    -(1/2)\alpha e^{-4\varphi}\widetilde{\Gamma}^m{}_{k \ell}
    \widetilde{\gamma}^{k\ell}(\partial_j\widetilde{\gamma}_{m i})
    \nonumber\\
    &\qquad
    +(1/2)\alpha e^{-4\varphi}\widetilde{\gamma}^{m k}(\partial_k\partial_j
    \widetilde{\gamma}_{m i}) 
    +(1/2)\alpha e^{-4\varphi}\widetilde{\gamma}^{-2}(\partial_i\mathcal{S})
    (\partial_j\mathcal{S})
    -(1/4)\alpha e^{-4\varphi}(\partial_i \widetilde{\gamma}_{k\ell})
    (\partial_j\widetilde{\gamma}^{k\ell})
    \nonumber\\
    &\qquad
    +\alpha e^{-4\varphi}\widetilde{\gamma}^{k m}(\partial_k\varphi)
    \widetilde{\gamma}_{j i}\partial_m
    +\alpha e^{-4\varphi}(\partial_j\varphi)\partial_i
    -(1/2)\alpha e^{-4\varphi}\widetilde{\Gamma}^m{}_{k\ell}
    \widetilde{\gamma}^{k\ell}\widetilde{\gamma}_{j i}\partial_m
    +\alpha e^{-4\varphi}\widetilde{\gamma}^{m k}
    \widetilde{\Gamma}_{i j k}\partial_m
    \nonumber\\
    &\qquad
    +(1/2)\alpha e^{-4\varphi}\widetilde{\gamma}^{\ell k}
    \widetilde{\gamma}_{j i}\partial_k\partial_\ell
    +(1/2)e^{-4\varphi}\widetilde{\gamma}^{m k}(\partial_j
    \widetilde{\gamma}_{i m})(\partial_k\alpha)
    +(1/2)e^{-4\varphi}(\partial_j\alpha)\partial_i
    +(1/2)e^{-4\varphi}\widetilde{\gamma}^{m k}\widetilde{\gamma}_{j i}
    (\partial_k \alpha)\partial_m]\mathcal{G}^j
    \nonumber\\
    &\qquad
  +[-\widetilde{A}^k{}_i(\partial_k\alpha)
    +(1/9)(\alpha_j)K
    +(4/9)\alpha (\partial_i K)
    +(1/9)\alpha K\partial_i 
    -\alpha \widetilde{A}^k{}_i\partial_k
  ]\mathcal{A},\label{eq:CPM}\\
  \partial_t\mathcal{G}^i
  &=2\alpha \widetilde{\gamma}^{i j}\mathcal{M}_j
  +[4\alpha\widetilde{\gamma}^{ij}(\widetilde{D}_j\varphi)
    -\alpha\widetilde{\gamma}^{i j}\partial_j
    -(\partial_k\alpha)\widetilde{\gamma}^{i k}
  ]\mathcal{A},\label{eq:CPG}
  \\
  \partial_t\mathcal{A}
  &=
  \alpha K\mathcal{A},
  \label{eq:CPA}\\
  \partial_t\mathcal{S}
  &=  -2\alpha \widetilde{\gamma}\mathcal{A}.
  \label{eq:CPS}
\end{align}
\end{widetext}



\end{document}